\documentstyle[11pt,aaspp4,psfig]{article}

\begin{document}

\def\mathnew{\mathsurround=0pt}
\def\simov#1#2{\lower .5pt\vbox{\baselineskip0pt \lineskip-.5pt
        \ialign{$\mathnew#1\hfil##\hfil$\crcr#2\crcr\sim\crcr}}}
\def\simg{\mathrel{\mathpalette\simov >}}
\def\siml{\mathrel{\mathpalette\simov <}}
\def\Mesz{M\'esz\'aros~}
\def\beq{\begin{equation}}
\def\enq{\end{equation}}
\def\bea{\begin{eqnarray}}
\def\ena{\end{eqnarray}}
\def\bec{\begin{center}}
\def\enc{\end{center}}
\def\hl{\hline}
\def\tl{\tableline}
\def\lbr{\linebreak\noindent}
\def\half{{1 \over 2}}

\received{ 12/18/97 }
\accepted{}
\slugcomment{ApJL, subm 12/18/97}
\lefthead{ Rees \& \Mesz }
\righthead{ Refreshed Shocks and Afterglow Longevity}
\tighten

\title{ REFRESHED SHOCKS \\ and AFTERGLOW LONGEVITY in GRB}

\author{M.J. Rees\altaffilmark{1} \& P. \Mesz\altaffilmark{2} }

\altaffiltext{1}{Institute of Astronomy, Cambridge University, Madingley Road,
Cambridge CB3 0HA, U.K.}

\altaffiltext{2}{Dpt. of Astronomy \& Astrophysics, Pennsylvania State
University, University Park, PA 16803}



\begin{abstract}

We consider fireball models where the ejecta have a range of bulk Lorentz
factors, so that the inner (lower $\Gamma$) parts may carry most of the mass, or
even most of the energy. The outer shock and contact discontinuity decelerate as
the fireball sweeps up external matter. This deceleration allows slower ejecta
to catch up, replenishing and reenergizing the reverse shock and boosting the
momentum in the blast wave. In consequence, the energy available to power the
afterglow may substantially exceed that of the burst itself.  Such models allow
a wide range of possibilities for the afterglow evolution, even in the case of
spherically symmetric expansion.

\end{abstract}

\keywords{gamma-rays: bursts}


\section{Introduction}
\label{sec:intro}

Afterglows from Gamma-Ray Bursts (GRB) have been discovered in at least
seven objects at X-rays wavelengths. Two of these also yielded optical
afterglows and one was followed in up to the radio band (e.g. papers
at the 4th Huntsville GRB Symposium, Meegan et al, 1997). The first
afterglow discovered at both X-ray and optical wavelengths, GRB 970228
(Costa et al, 1997) strongly favored a cosmological origin, and this was
strengthened even further by the discovery in the second object, GRB 970508,
of several systems of absorption lines yielding a limit $0.835 \siml z
\siml 2.3$ (Metzger et al, 1997). The power law flux decay predicted from
the simplest cosmological fireball afterglow model (\Mesz \& Rees, 1997)
turned out to be in good general agreement with the observed behavior (e.g.
\cite{tav97};\cite{wax97a};\cite{vie97b};\cite{wrm97};\cite{rei97}).
The simple model is also successful in explaining the rise and decay of
the light curves (\cite{wrm97}), as well as the overall  
radio behavior (\cite{good97}; \cite{wax97c}).
The ensuing massive observational campaigns on these sources
have, in the meantime, provided such excellent data that it is now feasible to
investigate  more subtle effects. Among these are, e.g. possible oscillations
superposed on the power law
decay (e.g. \cite{gal97}), the appearance of optical and X-ray bumps after
$\sim 1.5$ days on the optical and X-ray light curves (\cite{piro97}), the
evidence for an initial slow optical decay before the onset of the power
law rise and decay (Pedersen, et al, 1997), etc.  Most and perhaps all of
these features may be understood within the context of anisotropic outflows, 
or of isotropic outflows in inhomogeneous media (\cite{mrw97}).
It remains, however, an interesting question to investigate how slow can
a power law decay behavior be in a simple spherically symmetric model,
and what fraction of the total energy budget can be associated with the
late afterglow, as opposed to the initial GRB. 
 In most earlier discussions, the ejecta have been treated as a uniform sphere
or shell, where most of the energy
is carried by material with a well-defined (and generally high) Lorentz factor.
But it is possible -- perhaps, indeed, more physically realistic -- that  the
ejecta consist,  in effect, of many concentric  shells moving at different
(relativistic) speeds. This could be the outcome, for instance, of relativistic
shock propagation down a density gradient; or the inner shells might be more
heavily loaded with baryonic ejecta.  A given shell would then expand freely
until the contact
discontinuity had been decelerated, by sweeping up external material, to a
Lorentz factor lower than its own. It would then crash into a reverse shock,
thermalizing its energy and boosting the power of the afterglow.
   In this paper, we illustrate this phenomenon by discussing simple cases. We
show that such models can produce very slow decay rates in the afterglow; they
can accommodate very large (as well as small) ratios of afterglow to burst
fluences, and can naturally explain a variety of afterglow light curves.

\section{Kinematics of Ejecta with a Power-Law Radial Profile}
\label{sec:kin}

    The 'trigger' for a gamma ray burst may have complex time structure spread
over the duration (typically 10 seconds) of the intense gamma ray emission
itself. However, when we are considering mechanisms for the (much more
prolonged)  afterglow, we can treat this energy release as impulsive. It gives
rise to relativistically expanding debris  which eventually slows down as it
sweeps up external matter.  If the debris were spherical, it would be
characterized by the way its energy and mass were distributed among shells of
different Lorentz factor. The actual '$\Gamma$-profile' would be the outcome of
complex dynamics during the burst itself: shock fronts propagating down density
gradients, internal shocks, etc. A range of Lorentz factors seems likely --
indeed it is not obvious whether the energy (or even the mass) would be
concentrated towards the high or low end of the $\Gamma$ distribution. This is
our motivation for first  exploring a range of power-law' models, which can be
treated analytically.
   In simple cases where the expansion is ultrarelativistic (bulk Lorentz factor
$\Gamma \gg 1$) and all physical variables depend as power laws on radius (or
source-frame time), the dynamics can be treated in a 1-D approximation, which
will be valid either for spherical expansion or for jet angles $\simg
\Gamma^{-1}$. We consider the evolution at large radii, after the input of
mass and energy from the central source has ceased. In the source frame,
after deceleration by the external medium becomes important, the Lorentz
factor of the outer shock and the contact discontinuity begin to decrease as
\beq
\Gamma_c \propto r^{-n},
\enq
When the ejecta have, in effect, a single (high) Lorentz factor, we have the
well-known models where $n=3/2$ for adiabatic (weak coupling), or $n=3$ for 
radiative (strong coupling) expansion in a homogeneous external medium 
(for an inhomogeneous medium, see \Mesz, Rees \& Wijers, 1997). 
In the cases treated here, where the ejecta has a range of Lorentz factors  
$\Gamma_f$, the expansion will still obey a power law, but with different $n$.
At increasing $r$ the contact discontinuity, if it expands according to (1),
lags behind the light cone by a
fractional amount
\beq
{\Delta r \over r} \propto {1 \over 2 r} \int \Gamma_c^{-2}(r)~dr
\propto ~{1 \over 2 r} \int^r r^{-2n}~dr
\propto {1 \over 2(2n+1)}~\Gamma_c^{-2}(r)~.
\enq
We assume that the input of energy is variable, but occurs over a finite time
which is short compared to the deceleration time of the contact discontinuity,
so the impulsive approximation is valid. Subsequent shells of fireball
material endowed with a decreasing Lorentz factor $\Gamma_f$, catch up with
the contact discontinuity at radii $r$ which satisfy
$(\Delta r / r)= [2(2n+1)]^{-1} \Gamma_c^{-2}= (1/2) \Gamma_f^{-2}$,
or
\beq
\Gamma_f(r) = (2n+1)^{1/2} \Gamma_c(r)~,
\enq
where we have neglected the thickness of the shell of reverse shocked gas
relative to $r$. This is shown in Figure 1, where $r$ is distance from the
center of the outflow
and $t$ is source-frame (not detector) time, $t=r/c$. The impacting material
therefore has a Lorentz factor which is $(2n+1)^{1/2}$ times larger than
that of the contact discontinuity. (The latter has been decelerated, whereas
the former, having moved at constant speed, reaches the same radius with a
higher speed, see Figure 1). The ratio of the Lorentz factors, however, is only
a factor $\sim 2$, so the reverse shock(s) are marginally relativistic at all
times.  It then makes little  difference to the overall
dynamics whether the shock is radiative or adiabatic.

\section{Effects of a $\Gamma_f$-dependent  Mass and Energy Input}
\label{sec:dyn}

We assume that the primary event leading to the GRB produces a fireball
in which the mass fraction ejected above a given initial bulk Lorentz
factor decreases according to
\beq
M(\geq \Gamma_f ) \propto \Gamma_f^{-s}~.
\label{eq:mlordep}
\enq
The mass is dominated by the low $\Gamma_f$ shells for $s >0$, while the
energy will also be dominated by low $\Gamma_f$ material if $s >1$.
The mass that has 'caught up' with the contact discontinuity, and passed through
the reverse shock,  before it gets out to radius $r$ is (assuming $\Delta r/r
\ll 1$)
\beq
M(<r) \propto r^{n s}
\enq
The momentum impacting on the contact discontinuity from the reverse shocked
gas, per unit solid angle and measured in the comoving frame, is
\beq
 p_{rev} \propto {d M(<r ) \over dr} \propto r^{n s -1} ~.
\label{eq:revpres}
\enq
The impact is always marginally relativistic, $\Gamma_{rev} \sim 2 \sim$
constant. The force on the other (forward shock) side of the contact
discontinuity is
\beq
p_{for} \propto \rho_{ext} \Gamma_c^{1+A} r^2~,
\label{eq:forpres}
\enq
where $A=0$ for radiative (strong coupling) and $A=1$ for adiabatic (weak
coupling) expansion of the remnant (\Mesz, Rees \& Wijers, 1997), and the 
geometrical factor $r^2$ is included because this is
per unit solid angle. Here the adiabatic or radiative dependence is important,
since the outer shock is ultrarelativistic. We note that $A=0$ and $A=1$ are
extremes of the remnant behavior; intermediate values are also expected in 
realistic cases, e.g., as shown in the numerical calculations of 
Panaitescu \& \Mesz, 1997.  For $\rho_{ext}$ independent of
radius, equating the powers of $r$ in equations (\ref{eq:revpres}) and
(\ref{eq:forpres}) gives $n s -1 = -n(1+A) +2$, so the index $n = 3/(s + 1+A)$ 
and the radius dependence of the contact discontinuity Lorentz factor in the 
presence of a continuous input of mass and energy is
\beq
\Gamma_c \propto r^{-n} \propto r^{-3/(s+1+A)}~.
\label{eq:Gammac}
\enq
The value of this index is shown in Table 1 for various cases. 

We see that, for $s=0$ (where the later arriving, slower mass makes only
a logarithmic difference to the total), we recover as expected the previous
law of $n= 3/2 (3)$ for the $A=0(1)$ cases corresponding to the adiabatic 
(radiative) dynamic regimes. However for $s=1$, where most of the mass
is concentrated at lower $\Gamma_f$ we get $n=1(3/2)$ for $A=0(1)$; and for
$s=3/2$, where also most of the energy is dominated by the low $\Gamma_f$,
we get $n=6/7 (6/5)$ for $A=0(1)$.

\section{ Luminosity and Flux Evolution}
\label{sec:lum}

For $s > 1$ the total energy available for the afterglow gradually increases
during the deceleration,  but clearly the power law
dependences will only be valid over a finite range of times. Keeping these
limitations in mind, the luminosity dependence at late times can be evaluated.
For a contact discontinuity evolution given by equation (\ref{eq:Gammac}),
the energy liberated by the shock is $E \propto \Gamma_c^{1-s} \propto 
r^{-3(1-s)/(s+1+A)}$ and the observer time over which this is released is 
$T \propto r/\Gamma_c^2 \propto r^{(s+A+7)/(s+1+A)}$, corresponding to 
$r \propto T^{(s+1+A)/(s+A+7)}$. The ``kinetic" luminosity or flux is then
\beq
F_{kin} \propto r^{-(10-2s+A)/(s+1+A)}~\propto T^{-(10-2s+A)/(7+s+A)}~.
\label{eq:fkin}
\enq
where $T$ is detector (observer-frame) time. E.g., for $s=3/2$, $A=1$ we have
$F_{kin} \propto T^{-16/19}$.

If the radiation comes from the forward shock, and is due to synchrotron
radiation of shock-accelerated power law electrons in a turbulently
generated magnetic field which has built up to some fraction of equipartition
with the protons, we have a comoving synchrotron peak frequency $\nu'_m 
\propto \gamma_e^2 B' \propto \Gamma_c^3 \propto r^{-9/(s+1+A)} \propto 
T^{-9/(s+A+7)}$, and the corresponding observer-frame value is $\nu_m \propto
\Gamma_c \nu'_m \propto T^{-12/(s+A+7)}$, where $\gamma_e$ is the electron 
random Lorentz factor. Following \cite{mrw97} the comoving synchrotron peak
intensity is $I'_{\nu'_m} \propto n'_e \gamma_e {\nu'}_m^{-1} e_{sy}$, where
$n'_e \propto \Gamma_c n_{ext} \sim \Gamma_c$ is the comoving postshock 
electron density for expansion in a homogeneous external medium,
and $e_{sy} \propto (t'_{dyn}/t'_{sy})^a \propto r^a \Gamma_c^{2a}$ is
the synchrotron radiative efficiency of the electrons responsible for 
radiating at the peak frequency. For $a=0$ the electrons are radiatively
efficient and $e_{sy}=1$, while for $a=1$ the electrons are adiabatic 
(radiatively inefficient) and $e_{sy}<1$. Thus $I'_{\nu'_m} \propto
r^a \Gamma_c^{2a-1} \propto T^{[a(s+1+A)-6a+3]/(s+A+7)}$. The observer frame
peak synchrotron flux is
\beq
F_{\nu_m} \propto T^2 \Gamma_c^5 I'_{\nu'_m} \propto
 T^{-[6a-(s+1+A)(2+a)]/(s+A+7)}
\label{eq:fnum}
\enq
For a synchrotron intensity spectral slope $\beta$ above the synchrotron 
peak, $I_\nu \propto \nu^{\beta}$, the time decay of the flux in a
fixed detector frequency band $\nu_D$ (e.g. optical) is
\beq
F_D ~\propto ~ F_{\nu_m} (\nu_D/\nu_m)^{\beta}
\propto T^{[(2+a)(s+1+A)-6a+12\beta ]/(s+A+7)}
\label{eq:fdfor}
\enq
For $s=3/2,~A=1$ this is $\propto T^{[7-(5a/2)+12\beta]/(19/2)}$ and for 
$a=1,~\beta=1/3$ (e.g. synchrotron radiation below the peak from adiabatic
electrons) the flux would increase $\propto T^{17/19}$, steeper than in our 
previous models. For $a=1,~\beta=-1$ (above the peak) the flux is $\propto 
T^{-15/19}$. Similar considerations can be made for the reverse shock; however, 
if Rayleigh-Taylor instabilities occur across the contact discontinuity (e.g. 
Waxman \& Piran, 1994), these would tend to equalize the magnetic and nonthermal
energy densities, leading to a similar time-dependence of the emission.
Further complications may arise if the electrons responsible for $\nu_m$ are
adiabatic but those responsible for $\nu_D$ are radiative, e.g. equation (9) 
in \cite{mrw97} and related discussion (also \cite{spn97}).
Note that equation (\ref{eq:fdfor}) above was calculated
assuming $\beta \siml -1$. For $\beta \simg -1$ most of the energy does not
reside at $\nu_m$ but at some largest frequency $\nu_{cut}$ to which the
slope $\beta$ extends to, and the equation replacing (\ref{eq:fdfor})
would depend on this cutoff frequency. The cutoff can be computed, given model 
hypotheses defining the absolute values of the magnetic field and $\Gamma_c$.
Note also that equations (\ref{eq:Gammac},\ref{eq:fkin},\ref{eq:fnum},
\ref{eq:fdfor}) can be easily generalized to take into account a dependence on 
an external medium that depends on radius as $n_{ext} \propto r^{-d}$, as in 
\cite{mrw97}, leading to flatter decay laws than in equation ({\ref{eq:fdfor}).
In Table 1 we show the values of the time exponents of the fluxes in a fixed
detector band for a homogeneous external medium and some representative values 
of $s,~A,~a$ and $\beta$.

\section{Discussion}
\label{sec:disc}

We have considered here GRB models where the fireball does not constitute a
single ultrarelativistic shell (or uniform sphere), but contains material with a
range of speeds, larger
amounts of mass and energy being  characterized by lower  bulk
Lorentz factors. After the central energy-generation  has ceased the impulsive
approximation is valid, and during its evolution the later portions of the
ejecta catch up
with the previously decelerated matter, thus continually
re-energizing the shock that provides the afterglow of the GRB. We have shown
in \S \ref{sec:lum} that, depending on the Lorentz factor dependence of the
mass and energy injection, even spherically symmetric outflows can produce
very flat flux time decay laws (equations [\ref{eq:fdfor}]).
Even flatter decay laws are possible for synchrotron energy spectral indices
$\beta \simg -1$. A significant departure from our previous versions of an
isotropic GRB afterglow model is that here the flux decay law is not
dependent only on the spectral index (or on the density or angular dependence 
indices, c.f.  \Mesz, Rees \& Wijers, 1997) but also on the index $s$ of 
the Lorentz factor dependence of the mass and energy ejection (equation 
[\ref{eq:mlordep}]). Even for a homogeneous medium and for isotropic outflow, 
plausible values of $s$ can lead to afterglow luminosities whose 
time-integrated value can either exceed or fall below the initial $\gamma$-ray 
fluence. The former case would represent a situation similar to the 
hypernova model of Paczy\' nski, 1997.

Although we have focused on power-law dependences (cf equation (1) and (4)), 
our considerations clearly show that even  a spherical model could generate 
more complicated light curves. For example, suppose that, rather than 
containing material with a smooth distribution of   $\Gamma_f$, the fireball 
could be better modeled by two shells with very different $\Gamma_f$, the 
slower shell carrying more energy. The afterglow luminosity would initially 
decline as in the standard mono-energetic case; but when the blast wave had 
slowed down enough for the slower shell to catch up, there would be a boost 
to the luminosity before it again resumed a power-law decline. Changes in the 
slope of the flux decay laws (equations [\ref{eq:fdfor}]),or  
late bumps (e.g. Piro et al, 1997), could thus  be readily incorporated even 
into spherical models. The ``internal engine" associated with a catastrophic 
disruption or collapse event would inevitable lead to non-sphericity in the 
outflow; if the ejecta consisted of discrete overdense blobs, their 
deceleration could lead to more complex time-structure, or apparent 
shorter-term variability  (Galama et al, 1997, Pedersen, et al, 1997)
superposed on the overall power law behavior.

\acknowledgements{This research has been supported by the Royal Society,
NASA NAG5-2857 and NATO CRG-931446. We are grateful to A. Panaitescu and
R.A.M.J. Wijers for valuable comments.}


\begin{table*}
\begin{center}
\begin{tabular}{rrrrrrrr}
\tl \tl
   &    &     &   & $F_{kin}$ & $F_{D}$ & $F_{D}$ & $F_{D}$  \cr
\tl
 A & s  & n   & a &       & $\beta=1/3$ & $\beta=-1$ & $\beta=-3/2$  \cr
\tl
 0 & 0  &3/2  & 0 & -10/7 & 6/7   & -10/7 & -16/7  \cr
 0 & 1  & 1   & 0 & -1    & 1     &  -1   & -14/8  \cr
 0 & 3/2& 6/7 & 0 & -14/17& 18/17 & -14/17& -26/17 \cr
 1 & 0  & 3   & 0 & -11/8 & 1     & -1    & -14/8  \cr
 1 & 1  & 3/2 & 0 & -1    & 10/9  & -2/3  & -12/9  \cr
 1 &3/2 & 6/5 & 0 & -16/19& 22/19 & -10/19& -22/19 \cr
 1 & 0  & 3   & 1 & -11/8 &  1/2   & -12/8  & -18/8  \cr
 1 & 1  & 3/2 & 1 & -1    &  7/9   & -1     & -15/9 \cr
 1 &3/2 & 6/5 & 1 & -16/19& 17/19 & -15/19 & -27/19 \cr
\tl
\end{tabular}
 
\tablenum{1}
\caption{ For strong coupling (``radiative", $A=0$) or weak coupling 
(``adiabatic", $A=1$) remnant evolution in a homogeneous external medium and 
a mass outflow index $s$ 
($M(\geq \Gamma_f ) \propto \Gamma_f^{-s}$) the bulk Lorentz of the contact 
discontinuity evolves as $\Gamma_c \propto r^{-n}$, producing a kinetic flux 
$F_{kin} \propto t^v$ whose index $v$ is in column 5. For different spectral 
indices $\beta$ the forward shock produces a flux in a fixed detector band 
$F_{D} \propto t^w$, whose indices $w$ are shown in columns 6 through 8 for 
radiative ($a=0$) or adiabatic ($a=1$) electrons. Flatter decay laws would be 
obtained if the external density decreases with radius.} 
\end{center}
\end{table*}
 

\begin{figure}
\centerline{\psfig{figure=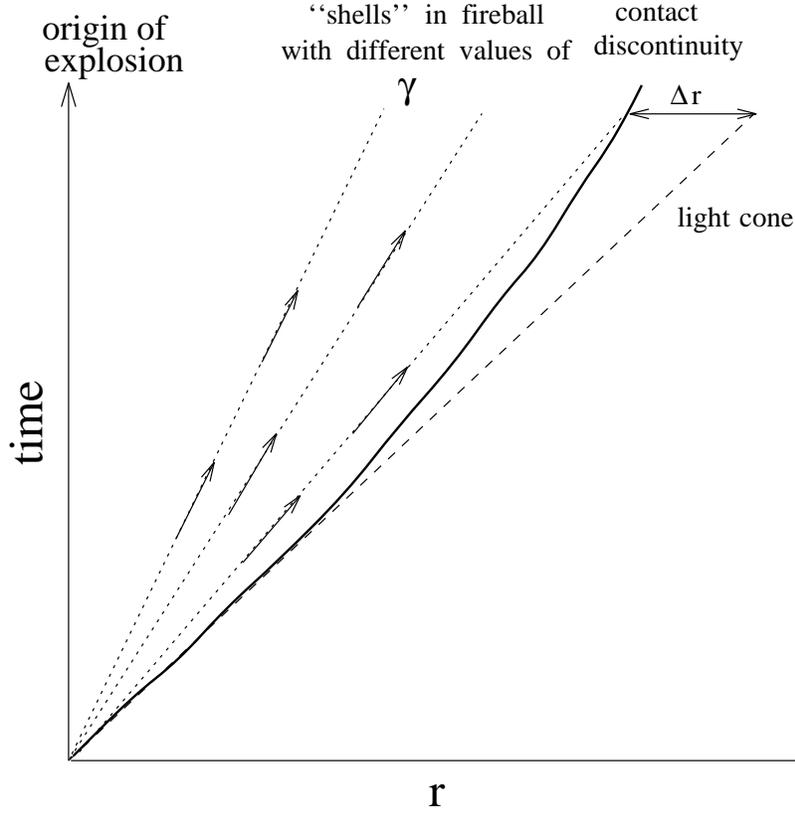}}
\vspace*{1in}
\figcaption{Schematic space-time diagram in source frame coordinates
of a  relativistic outflow, with a range of Lorentz factors $\Gamma_f$, 
triggered by an explosion that can be approximated as instantaneous
(decreasing $\Gamma_f$ values lead to world lines further to the left,
which are straight before entering the shock).
The contact discontinuity and the forward shock are being decelerated due to 
the increasing amount of external matter being swept up, so that they lag 
behind the light cone by an increasing amount amount $\Delta r$ whose 
increase with $r$ is steeper than linear. This deceleration allows slower 
ejecta (with lower $\Gamma_f$) to catch up and pass through a  reverse 
shock just inside the contact discontinuity.   Both shocks  are here  
approximated as having negligible width.  Energy and momentum from 
progressively slower material is thus continuously  re-energizing the 
shocks, resulting in a more gradual decay of the afterglow.}
\end{figure}


\begin{thebibliography}{}

\bibitem[Costa et al, 1997a]{cos97a} Costa, E, et al., 1997s, Nature, 387, 783
\bibitem[Galama, et al., 1997]{gal97} Galama, T., et al., 1997, Nature 387, 479
\bibitem[Goodman, J., 1997]{good97} Goodman, J., 1997, New Astronomy, 2(5), 449
\bibitem[Meegan et al., 1997]{mee97} Meegan, C, Preece, R and Koshut, T, 1997, 
 eds, {\it 4th Huntsville GRB Symposium}, (New York: AIP), in press.
\bibitem[\Mesz \& Rees, 1997a]{mr97a} \Mesz, P \& Rees, M.J., 1997a, \apj, 476,
 232
\bibitem[\Mesz, Rees \& Wijers 1997]{mrw97} \Mesz, P, Rees, M.J. \& Wijers, 
 R.A.M.J., 1997, \apj, in press (astro-ph/9709273)
\bibitem[Metzger et al., 1997]{metz97} Metzger, M et al., 1997, Nature, 387, 878
\bibitem[Paczy\'nski, 1997]{pac97} Paczy\'nski, B., 1997, \apj, in press
 (astro-ph/9710086)
\bibitem[Panaitescu \& \Mesz, 1997]{panm97} Panaitescu, A. \& \Mesz, P, 1997, 
 \apj, subm (astro-ph/9711339)
 (astro-ph/9710086)
\bibitem[Pedersen, et al, 1997]{ped97} Pedersen, H., 1997, preprint
 (astro-ph/9710322)
\bibitem[Piro, et al, 1997]{piro97} Piro, L., et al, 1997, preprint
 (astro-ph/9710355)
\bibitem[Reichart, 1997]{rei97} Reichart, D., 1997, \apjl, in press
 (astro-ph/9704198)
\bibitem[Sari, Piran \& Narayan, 1997]{spn97} Sari, R., Piran, T. \& Narayan, R,
 1997, \apjl, subm (astro-ph/9712005)
\bibitem[Tavani, 1997]{tav97} Tavani, M., 1997, \apjl, 483, L87
\bibitem[Vietri, 1997b]{vie97b} Vietri, M., 1997b, \apjl, subm
 (astro-ph/9706060)
\bibitem[Waxman \& Piran, 1994]{wapi94} Waxman, E. \& Piran, T., 1994, \apjl, 
 433, L85
\bibitem[Waxman, 1997a]{wax97a} Waxman, E., 1997a, \apjl, subm
(astro-ph/9704116)
\bibitem[Waxman, et al, 1997c]{wax97c} Waxman, E., Kulkarni, S. \& Frail, D.,
  1997c, \apj, subm (astro-ph/9709199)
\bibitem[Wijers, Rees \& \Mesz, 1997]{wrm97} Wijers, R.A.M.J., Rees, M.J. \&
  \Mesz, P., 1997, \mnras, 288, L51

\end{thebibliography}
\end{document}